\documentclass[aps,twocolumn,prl,groupedaddress,superscriptaddress,amsmath,amssymb,amsthm,showpacs]{revtex4}
\usepackage{subfigure,hyperref,bbm,times}
\usepackage[T1]{fontenc}
\usepackage{braket}
\usepackage{amsthm}
\usepackage{epsfig}
\usepackage{color}
\usepackage{graphicx}% Include figure files
\usepackage{dcolumn}% Align table columns on decimal point
\usepackage{bm}% bold math

\providecommand{\openone}{\leavevmode\hbox{\small1\kern-3.8pt\normalsize1}}

\usepackage{soul}

\begin{document}

\title{Indistinguishability-enabled coherence for quantum metrology}

\author{Alessia Castellini}
\email{alessia.castellini@unipa.it}
\affiliation{Dipartimento di Fisica e Chimica, Universit\`a di Palermo, via Archirafi 36, 90123 Palermo,
Italy}

\author{Rosario Lo Franco}
\email{rosario.lofranco@unipa.it}
\affiliation{Dipartimento di Fisica e Chimica, Universit\`a di Palermo, via Archirafi 36, 90123 Palermo, Italy}
\affiliation{Dipartimento di Ingegneria, Universit\`{a} di Palermo, Viale delle Scienze, Edificio 6, 90128 Palermo, Italy}

\author{Ludovico Lami}
\email{ludovico.lami@gmail.com}
\affiliation{School of Mathematical Sciences and Centre for the Mathematics and Theoretical Physics of Quantum Non-Equilibrium Systems, University of Nottingham, University Park, Nottingham NG/ 2RD, United Kingdom}
\email{ludovico.lami@gmail.com}

\author{Andreas Winter}
\email{andreas.winter@uab.cat}
\affiliation{ICREA \& F\'{i}sica Te\'{o}rica: Informaci\'{o} i Fen\'{o}mens Qu\`{a}ntics, Departament de F\'{i}sica, Universitat Aut\`{o}noma de Barcelona, ES-08193 Bellaterra
(Barcelona), Spain}

\author{Gerardo Adesso}
\email{Gerardo.Adesso@nottingham.ac.uk}
\affiliation{School of Mathematical Sciences and Centre for the Mathematics and Theoretical Physics of Quantum Non-Equilibrium Systems, University of Nottingham, University Park, Nottingham NG/ 2RD, United Kingdom}

\author{Giuseppe Compagno}
\email{giuseppe.compagno@unipa.it}
\affiliation{Dipartimento di Fisica e Chimica, Universit\`a di Palermo, via Archirafi 36, 90123 Palermo, Italy}

\date{\today }% It is always \today, today,

\begin{abstract}
Quantum coherence plays a fundamental and  operational role in different areas of physics.  A resource theory has been developed to characterize the coherence of distinguishable particles systems. Here we show that indistinguishability of identical particles is a source of coherence, even when they are independently prepared. In particular, under spatially local operations, states that are incoherent for distinguishable particles, can be coherent for indistinguishable particles under the same procedure. We present a phase discrimination protocol, in which we demonstrate the operational advantage of using two indistinguishable particles rather than distinguishable ones. The coherence due to the quantum indistinguishability significantly reduces the error probability of guessing the phase using the most general measurements.
%Coherence of quantum systems plays a fundamental and operational 
% role in different areas of physics, which has been recognized for a 
% long time. Recently, a resource theory of coherence has been developed 
% to characterize the coherence of single-particle systems and of systems
% composed of multiple (distinguishable) particles. 
% Here we define the coherence for a system composed of two identical
% particles with two-level internal degrees of freedom, in the framework 
% of spatially local operations and classical communication. 
% We show that the indistinguishability of identical particles is a 
% source of coherence: while initially incoherent states of distinguishable
% particles remain incoherent under local operations, the same procedure
% applied to indistinguishable particles let coherence emerge.
% Furthermore, we present a phase discrimination protocol, in which we
% can demonstrate the operational advantage of using two indistinguishable 
%particles, rather than distinguishable ones. Concretely, the coherence 
%due to the quantum indistinguishability significantly reduces the 
%error probability of guessing the phase using the most general 
%measurements.
\end{abstract}



\maketitle
\textit{Introduction. ---}
Quantum coherence concerns the possibility to create a quantum state as a superposition of two or more different configurations.
Recently, a resource theory of coherence has been proposed, in order to characterize, quantify and exploit coherence \cite{chitambar2016critical,baumgratz2014quantifying,napoli2016robustness,piani2016robustness,de2016genuine,winter2016operational,colloquiumAdesso2017}. Chosen a reference basis, diagonal and non diagonal states in that basis are called incoherent (free) states and coherent (resource) states, respectively. The free operations to which we have access, in 
 accordance with physical constraints imposed on the system, are the 
 so-called \emph{incoherent operations}. They cannot create coherence 
 starting from incoherent states.  It has been shown that coherence of a quantum state has an operational relevance in the implementation of phase discrimination protocols using distinguishable particles: the robustness of coherence quantifies not only the amount of coherence of a quantum state but also the advantage offered by its presence in a metrology protocol \cite{napoli2016robustness,pires2018coherence}. In the context of quantum thermodynamics, a link between the coherence of a quantum state and the extractable work has been identified \cite{korzekwa2016extraction}, and very recently the role of quantum coherence as a resource for the non-equilibrium entropy production has been pinpointed \cite{Paternostro2017}.

The resource theory of coherence is well defined for single-particle systems and for systems of multi-distinguishable particles.
On the other hand, recently, it has been shown that indistinguishability of identical particles confers on quantum systems original properties that can be used as sources of entanglement to implement quantum information processes, impossible to perform if particles are distinguishable \cite{bose2002indisting,PhysRevLett.88.187903,PhysRevA.68.052309,PlenioExtracting,PhysRevA.65.062305,omarIJQI,swapping2018Castellini,LoFrancoPRL}.

In this Letter, we generalize the resource theory
 of coherence to systems of identical particles, using a particle-based approach that only allows physical labels to address the 
 particles in the system \cite{LFCSciRep,compagno2018dealing}. Our
 aim is to show how indistinguishability may be a source of operational coherence. Clearly, and 
 in contrast to the case of distinguishable particles, to achieve this
 we have to consider at least two (identical) particles.

\textit{Coherence of identical particle states. ---} Let us consider, in the no-label formalism \cite{LFCSciRep,compagno2018dealing,lourencco2019entanglement}, two identical spins (that is particles with a two-level internal state space) in the mixed state
\begin{equation}\label{I1}
\rho^{\mathcal{I}}=\sum_{\sigma,\tau=\downarrow,\uparrow}p_{\sigma\tau}\ket{\psi \sigma,\psi' \tau}\bra{\psi \sigma,\psi' \tau},
\end{equation}
where $\mathcal{I}$ stands for "identical", $\psi$ and $\psi'$ represent the spatial degrees of freedom and $\sigma$ and $\tau$ are the pseudospins. The coefficients $p_{\sigma\tau}$ are such that $\mathrm{Tr}[\rho^{\mathcal{I}}]=1$. Each term $\ket{\psi \sigma,\psi' \tau}$ in the sum is the state
 of two identical particles, one of which is characterised by
 $\psi$ and $\sigma$, the other by $\psi'$
 and $\tau$. This state is a global object, which crucially cannot be written as a tensor product of single-particle states,
 i.e., $\ket{\psi \sigma,\psi' \tau} \neq\ket{\psi \sigma}\otimes\ket{\psi' \tau}$. 

To define the coherence of this state, an orthonormal basis has to be chosen.  Let us now make some considerations in order to opportunely choose a "preferred" basis, depending on the possible operations that can be performed for free on the system. A resource theory requires the identification of a physically and operationally meaningful set of free operations, which enable manipulation of resourceful states. Analogously to the case of distinguishable particles, both single-particle and two-particle operations on the system are allowed. Identical particles are physically not addressable individually \cite{ghirardi2002,tichyFort}, thus the concept of single-particle operation may seem ill-defined, requiring to differentiate them. Furthermore, as a consequence of the fact that the global Hilbert space is not a tensor product of two single-spin spaces, if a CNOT (free operation for distinguishable particles \cite{winter2016operational}) has to be performed, it is not possible to differentiate between the control and the target particle.

Following the procedure to identify and quantify the useful entanglement between identical particles \cite{LoFrancoPRL}, in order to characterize the quantum coherence we here adopt a framework based on sLOCC, which consists in the identification of specific separate spatial regions to make single-particle and two-particle local measurements.
In particular, we choose the basis $\mathcal{B}^{I}=\{\ket{\mathrm{L} \sigma,\mathrm{R} \tau}; \ \sigma,\tau=\downarrow,\uparrow \}$, where $\ket{\mathrm{L}}$ and $\ket{\mathrm{R}}$ are two states spatially localized in the separate regions $\mathcal{L}$ and $\mathcal{R}$, respectively. If $\rho^{\mathcal{I}}$ is diagonal in the chosen basis, 
%and so of the form
%\begin{equation}\label{INC}
%\rho^{(\mathcal{I})}_{\mathrm{inc}}=\sum_{\sigma,\tau=\downarrow,\uparrow}a_{\sigma\tau}\ket{\mathrm{L\ \sigma,R \ \tau}}\bra{ \mathrm{L \ \sigma,R} \ \tau},
%\end{equation}
it is incoherent, %Instead, if it is of the following form
%\begin{equation}\label{Ic}
%\rho^{(\mathcal{I})}_{\mathrm{LR}}=\sum_{\sigma\tau\sigma'\tau'}a_{\sigma\tau\sigma'\tau'}\mathrm{\ket{L\ \sigma,R \ \tau} \bra{L \ \sigma',R \ \tau'} },
%\end{equation}
otherwise we say that $\rho^{\mathcal{I}}$ is coherent. We point out that if $\ket{\psi}$ and $\ket{\psi'}$ spatially overlap, the measurement regions $\mathcal{L}$ and $\mathcal{R}$ are taken within the shared spatial region.

For non-identical particles, identified by labels A and B, the state of Eq.~\eqref{I1} would correspond to $\rho^{\mathcal{NI}}=\sum_{\sigma,\tau}p_{\sigma\tau}(\ket{\psi \sigma}\bra{\psi \sigma})_{\mathrm{A}}\otimes (\ket{\psi' \tau}\bra{\psi' \tau})_{\mathrm{B}}$. It can be shown that $\rho^{\mathcal{NI}}$ is incoherent in the basis $\mathcal{B}=\{\ket{\mathrm{L} \sigma}_{\mathrm{A}}\otimes \ket{\mathrm{R} \tau}_{\mathrm{B}}; \sigma,\tau=\downarrow,\uparrow \}$ (see Supplemental Material). Projecting, on the other hand, $\rho^{\mathcal{I}}$ of Eq.~\eqref{I1} on the chosen subspace, we obtain the following state
\begin{align}\label{IRL}
\begin{split}
\rho^{\mathcal{I}}_{\mathrm{LR}}=\dfrac{1}{\mathcal{N}}\sum_{\sigma, \tau=\downarrow,\uparrow}p_{\sigma\tau}&(|l|^2|r'|^2\ket{\mathrm{L}\sigma,\mathrm{R}\tau}\bra{\mathrm{L}\sigma,\mathrm{R}\tau}\\ &+\eta lr'l'^{\ast}r^{\ast}\ket{\mathrm{L}\sigma,\mathrm{R}\tau}\bra{\mathrm{L}\tau,\mathrm{R}\sigma}\\
&+\eta l' r l^{\ast} r'^{\ast}\ket{\mathrm{L}\tau,\mathrm{R}\sigma}\bra{\mathrm{L}\sigma,\mathrm{R}\tau}\\
&+\eta |l'|^2|r|^2\ket{\mathrm{L}\tau,\mathrm{R}\sigma}\bra{\mathrm{L}\tau,\mathrm{R}\sigma}),
\end{split}
\end{align}
where $\mathcal{N}$ is a global normalization factor, $l=\braket{L|\psi}$, $r'=\braket{R|\psi'}$, $l'=\braket{L|\psi'}$ and $r=\braket{R|\psi}$ are the probability amplitudes of finding one particle in the two spatially separate states $\ket{\mathrm{L}}$ and $\ket{\mathrm{R}}$ and $\eta$ is $1$ for bosons and $-1$ for fermions. Assuming, for instance,  $l,r'\neq 0$, the remaining probability amplitudes $l'$ and $r$ are zero if and only if $\ket{\psi}$ and $\ket{\psi'}$ do not spatially overlap  \footnote{Notice that such a constraint in the definition of spatial overlap is made here for the sake of simplicity. Certainly, a general definition of spatial overlap can be straightforwardly achieved by using projectors onto all the possible bound states in each region $\mathcal{L}$ and $\mathcal{R}$, which will be reported elsewhere.}. If $p_{\downarrow\uparrow}=p_{\uparrow\downarrow}=0$, the state $\rho^{\mathcal{I}}$ does not contain coherence, regardless of the identity of particles. Otherwise, the state is coherent if and only if the particles spatially overlap (are indistinguishable). If they do not overlap, the state is incoherent: non-overlapping identical particles act like non-identical ones.

\begin{figure}[!t]
\centering
\includegraphics[scale=0.4]{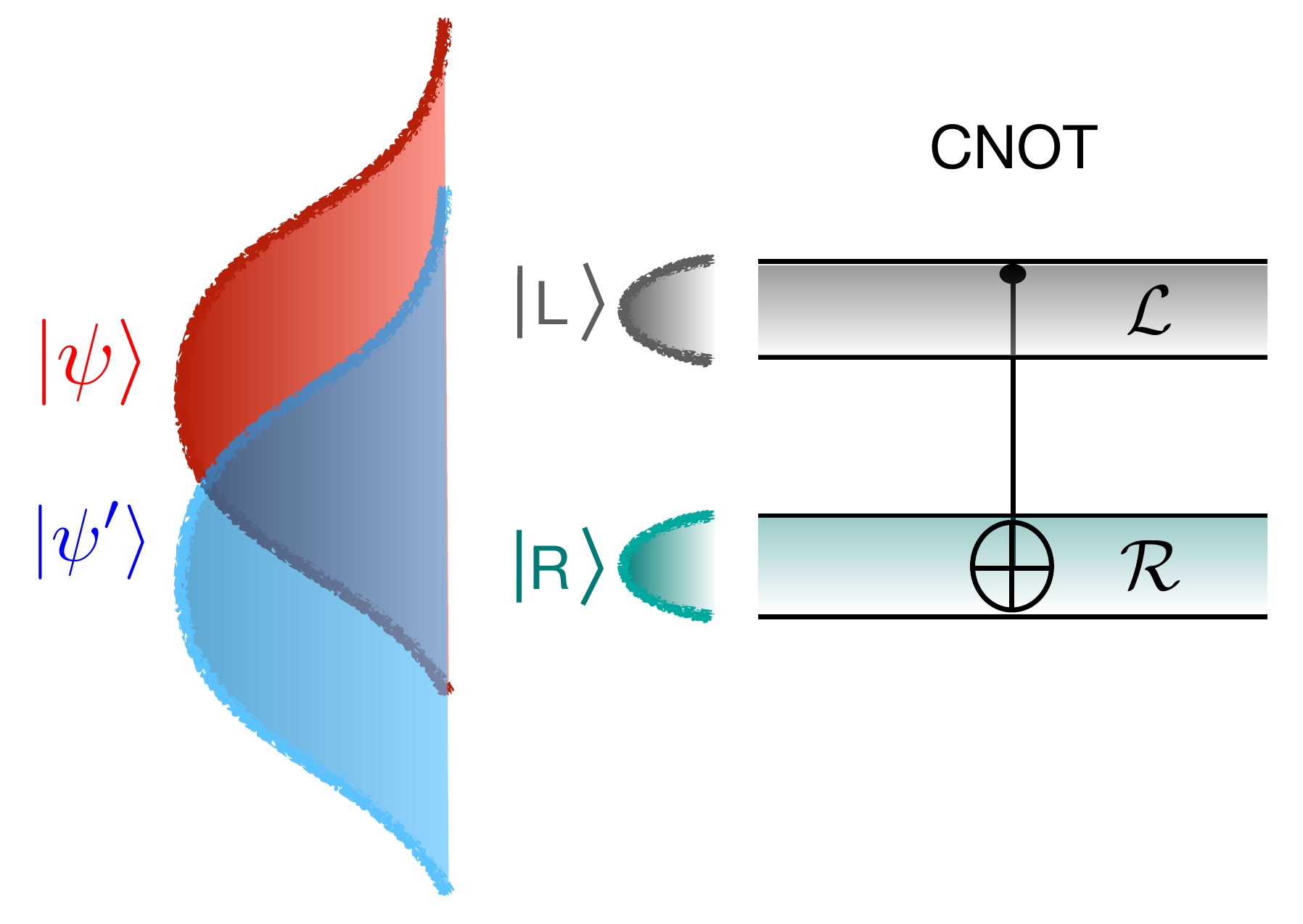}
\caption{CNOT gate for identical particles in the sLOCC framework.}
\label{CNOT}
\end{figure}

 %In this framework we define the incoherent states for two identical ($\mathcal{I}$) particles as the diagonal states in the operational basis
%\begin{equation}
%\rho_{\mathrm{inc}}^{(\mathcal{I})}=\sum_{\sigma\tau}p_{\sigma\tau}|\mathrm{L\sigma,R\tau\rangle\langle L\sigma,R\tau}|.
%\end{equation}
%In contrast with Eq.~(1),  $\rho_{\mathrm{inc}}^{(\mathcal{I})}\neq\sum_{\sigma\tau}p_{\sigma\tau}(|\mathrm{L\sigma \rangle \langle L\sigma|) \otimes (|R\tau\rangle\langle R\tau|).}$\\
%\indent Coherent states are of the form
%\begin{equation}
%|\Psi^{(\mathcal{I})}\rangle_{\mathrm{c}}=\sum_{\sigma\tau} \alpha_{\sigma\tau}|\mathrm{L\sigma,R\tau}\rangle.
%\end{equation}
\indent In the sLOCC framework, incoherent operations, known for non-identical particles, can be straightforwardly translated in the context of identical particle systems. Let us consider for example the CNOT gate. In the standard implementation with non-identical particles, there are individually addressed control and target spins.  By analogy, if we have identical particles, we can identify the measurement regions $\mathcal{L}$ and $\mathcal{R}$ as control and target regions, respectively (Fig. \ref{CNOT}). Applying it to the state of Eq.~\eqref{IRL} when it is incoherent, it can be easily shown that it gives us an incoherent state. Therefore the CNOT gate remains an incoherent operation which transforms incoherent states of identical particles into incoherent states. 

\textit{Phase discrimination protocol by particle indistinguishability. ---}
\begin{figure}[!b]
\centering
\includegraphics[scale=0.39]{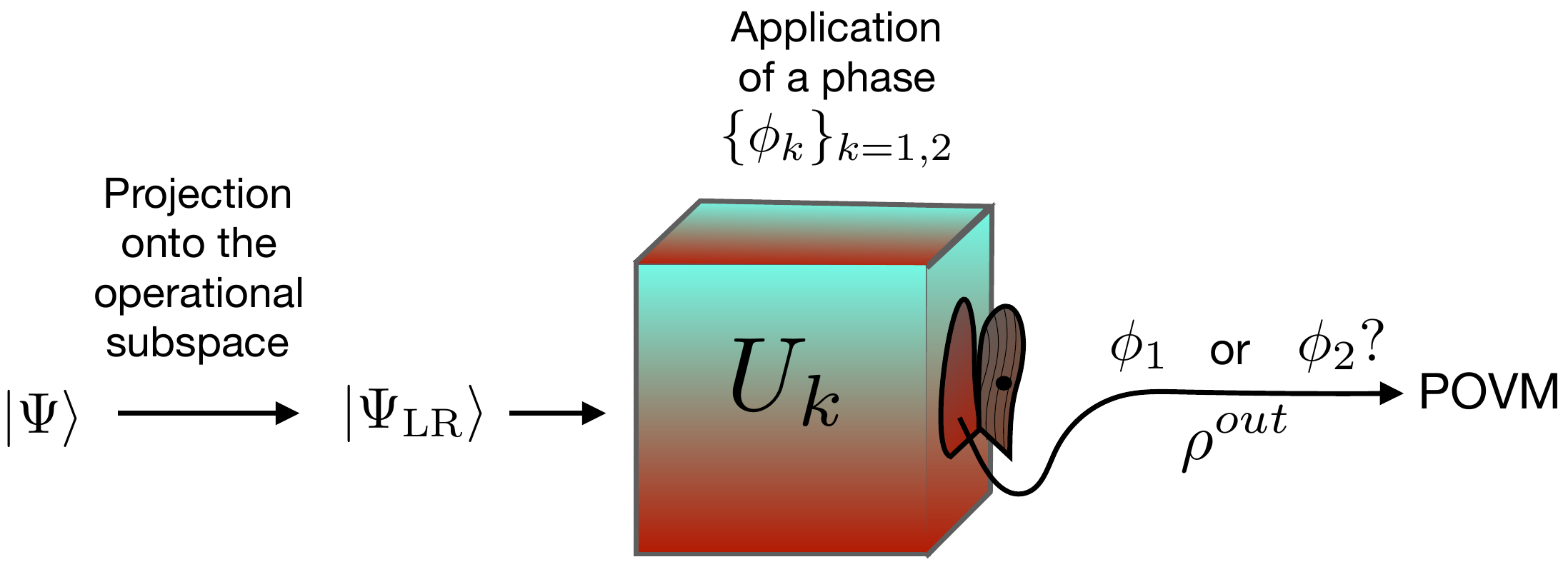}
\caption{Phase discrimination game.}
\label{Fig2}
\end{figure}
Our purpose is to determine when the contribution to  coherence due to the spatial overlap of identical spins plays an operational role in quantum information processing. It is known that within the context of quantum metrology, quantum coherence of states of non-identical particles is a resource for phase discrimination tasks \cite{napoli2016robustness,ringbauer2018certification}. We now describe an analogous game for indistinguishability-enabled quantum coherence.

\begin{figure*}[!t]
\includegraphics[scale=0.55]{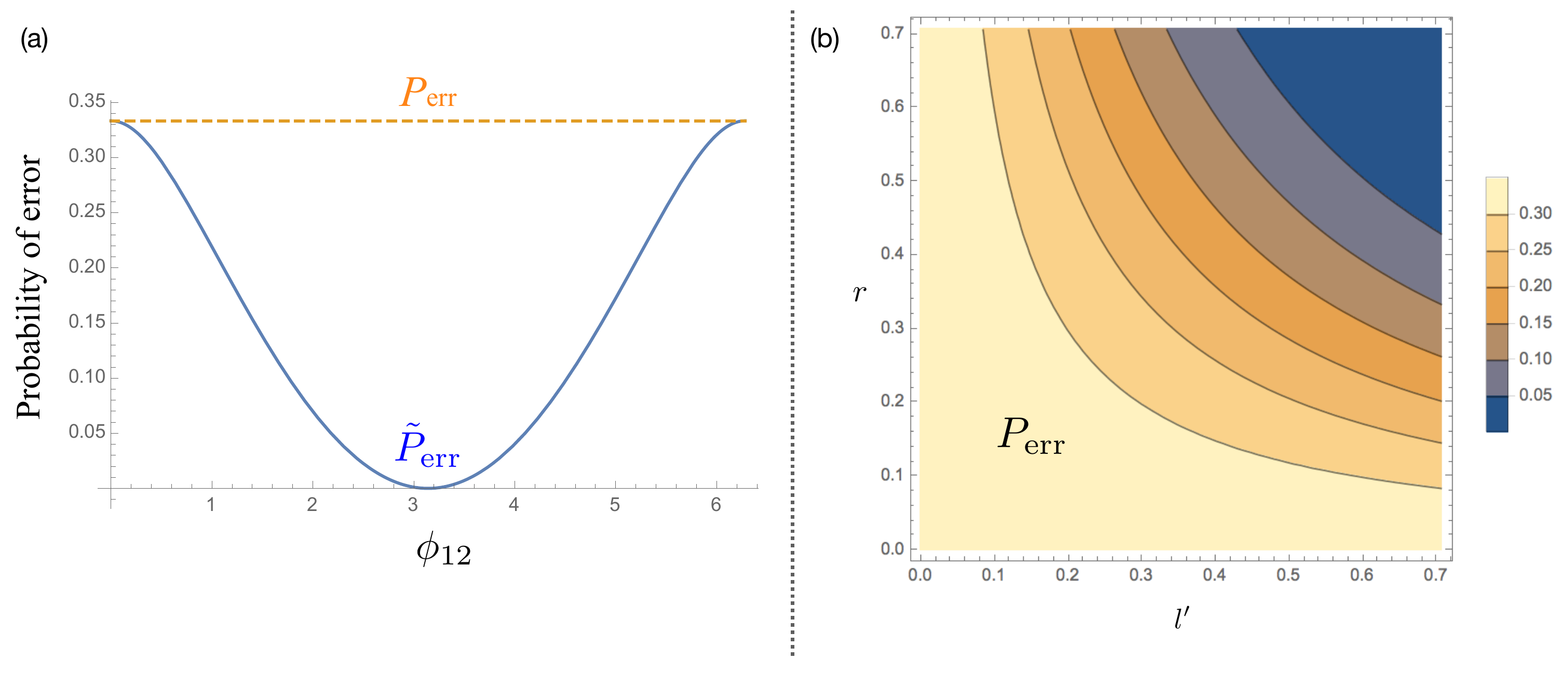}
\caption{{\small (a): Probability of error in function of the phase difference $\phi_{12}$, with $p_1=1/3$, $|l|^2=|r'|^2=\dfrac{1}{2}$, and $\omega_{\downarrow\uparrow}-\omega_{\uparrow\downarrow}=1$. The orange dashed line corresponds to $P_{\mathrm{err}}$ (Eq.~\eqref{Perr2}) when there is no overlap ($l'r=0$) or, analogously, to the case with distinguishable particles. The blue one corresponds to $\tilde{P}_{\mathrm{err}}$ (Eq.~\eqref{Perrover}), i.e. to the case of spatial overlap with $|l'|^2=|r|^2=\dfrac{1}{2}$. (b): Contour plot of $P_{\mathrm{err}}$ (Eq.~\eqref{Perr2}) in function of $r$ and $l'$, with $\phi_{12}=\pi$.}}
\label{Fig1}
\end{figure*}

The state of two independently-prepared non-identical spins, for instance, $\ket{\Psi}_{\mathrm{AB}}=\ket{\psi \downarrow_{\mathrm{A}},\psi'\uparrow_{\mathrm{B}}}=\ket{\psi \downarrow}_{\mathrm{A}}\otimes\ket{\psi'\uparrow}_{\mathrm{B}}$, is manifestly incoherent in the basis $\mathcal{B}$.
Differently, let us consider the corresponding pure state for two independently prepared identical spins
\begin{equation}\label{Psi}
|\Psi\rangle=|\psi \downarrow,\psi'\uparrow\rangle,
\end{equation}
whose projection in the operational subspace spanned by $\mathcal{B}^{\mathcal{I}}$ is
\begin{equation}\label{PsiRL}
\ket{\Psi_{\mathrm{LR}}}=\dfrac{1}{\mathcal{N}_{\mathrm{LR}}}(lr'|\mathrm{L}\downarrow,\mathrm{R}\uparrow\rangle+\eta l'r|\mathrm{L}\uparrow,\mathrm{R}\downarrow\rangle),
\end{equation}
with the normalization constant $\mathcal{N}_{\mathrm{LR}}=\sqrt{|lr'|^2+|l'r|^2}$.  A non-zero coherence for this state can only stem from indistinguishability ($l'r \neq 0$). Let us send $|\Psi_{\mathrm{LR}}\rangle$ in a box where a unitary transformation $U_k=e^{i\hat{G}\phi_k}$ ($\hat{G}$ is the generator of the transformation) applies one of the phases $\{\phi_k\}_{k=1,2}$ with a probability $p_k$ to the state (see Fig. \ref{Fig2}). The requirement is that $U_k$ is a free (incoherent) operation, so $\hat{G}=\sum_{\sigma\tau}\omega_{\sigma\tau}|\mathrm{L} \ \sigma, \mathrm{R} \ \tau\rangle \langle \mathrm{L} \ \sigma, \mathrm{R} \ \tau |$. The action of the box is
\begin{align}\label{Psik}
\begin{split}
|\Psi^k_{\mathrm{LR}}\rangle&=U_k |\Psi_{\mathrm{LR}}\rangle\\
&=\dfrac{1}{\mathcal{N}_{\mathrm{LR}}}[ lr' e^{i\omega_{\downarrow\uparrow}\phi_k}|\mathrm{L}\downarrow,\mathrm{R}\uparrow\rangle\\
& \hspace{1.2cm}+\eta l'r e^{i\omega_{\uparrow\downarrow}\phi_k}|\mathrm{L}\uparrow,\mathrm{R}\downarrow\rangle].
\end{split}
\end{align}
Due to our ignorance of which of the two phases has been applied, the state coming out of the box is in the following classical mixture
\begin{equation}\label{classmixt}
\rho^{out}=\sum_{k=1,2}p_k|\Psi^k_{\mathrm{LR}}\rangle\langle \Psi^k_{\mathrm{LR}}|.
\end{equation}
Physically, the effect of the box can be realized with $\hat{G}$ corresponding to the Hamiltonian of two independent spins subject to localized magnetic fields $\mathrm{B}_{\mathcal{L}}^k$ and $\mathrm{B}_{\mathcal{R}}^k$ which randomly occur with a probability $p_k$.

The objective now is to establish which phase has actually been applied. In other words, the phase discrimination game translates in a state discrimination one. The two states $\ket{\Psi^1_{\mathrm{LR}}}$ and $\ket{\Psi^2_{\mathrm{LR}}}$ are in general not orthogonal: in order to discriminate them, a positive operator-valued measure (POVM) \cite{Helstrom1976,audretsch} has to be chosen. We look for a POVM described by a set of operators $\{\hat{\Pi}_1,\hat{\Pi}_2\}$, each of which is associated to an outcome of the measurement: if we obtain 1, associated to the operator $\hat{\Pi}_1$, we conclude that the state coming of the box is $|\Psi^1_{\mathrm{LR}}\rangle$, if the result is 2, associated to the operator $\hat{\Pi}_2$, the state is $|\Psi^2_{\mathrm{LR}}\rangle$. The optimal POVM can be obtained by minimizing the probability of making an error in the discrimination, that is the probability of obtaining the result associated to $\hat{\Pi}_k$ when the state was $|\Psi_{\mathrm{LR}}^{k'}\rangle$ with $k'\neq k$. Such a probability is \cite{Helstrom1976,POVM}
\begin{equation}\label{Perr1}
P_{\mathrm{err}}=p_1\langle \Psi_{\mathrm{LR}}^1|\hat{\Pi}_2|\Psi_{\mathrm{LR}}^1\rangle+p_2\langle \Psi_{\mathrm{LR}}^2|\hat{\Pi}_1|\Psi_{\mathrm{LR}}^2\rangle.
\end{equation}
Using the property $ \sum_k\hat{\Pi}_k=\mathbb{I}$ \cite{audretsch}, we can rewrite Eq.~\eqref{Perr1} as follows 
\begin{equation}
P_{\mathrm{err}}=p_1-\mathrm{Tr}\left[(p_1|\Psi^1_{\mathrm{LR}}\rangle\langle \Psi^1_{\mathrm{LR}}|-p_2|\Psi^2_{\mathrm{LR}}\rangle\langle\Psi^2_{\mathrm{LR}}|)\hat{\Pi}_1\right].
\end{equation}
%\begin{align}
%\begin{split}
%P_{\mathrm{err}}&=p_1-p_1\langle \Psi^1_{\mathrm{LR}}|\hat{\Pi}_1|\Psi^1_{\mathrm{LR}}\rangle+p_2\langle \Psi^2_{\mathrm{LR}}|\hat{\Pi}_1|\Psi^2_{\mathrm{LR}}\rangle\\
%&=p_1-\mathrm{Tr}\left[(p_1|\Psi^1_{\mathrm{LR}}\rangle\langle \Psi^1_{\mathrm{LR}}|-p_2|\Psi^2_{\mathrm{LR}}\rangle\langle\Psi^2_{\mathrm{LR}}|)\hat{\Pi}_1\right].
%\end{split}
%\end{align}
The minimization of $P_{\mathrm{err}}$ is obtained when $\hat{\Pi}_1$ is the projector onto the positive eigenvector of the operator $\Delta=p_1|\Psi^1_{\mathrm{LR}}\rangle\langle \Psi^1_{\mathrm{LR}}|-p_2|\Psi^2_{\mathrm{LR}}\rangle\langle\Psi^2_{\mathrm{LR}}|$. Choosing a rotated orthonormal basis $|0\rangle$ and $|1\rangle$ (with $|0\rangle$ slicing in half the angle between $|\Psi^1_{\mathrm{LR}}\rangle$ and $|\Psi^2_{\mathrm{LR}}\rangle$), we can write
\begin{equation}
|\Psi^1_{\mathrm{LR}}\rangle=\cos \theta |0\rangle+\sin \theta |1\rangle, \ |\Psi^2_{\mathrm{LR}}\ \rangle=\cos \theta |0\rangle-\sin \theta |1\rangle.
\end{equation}
The positive eigenvalue of $\Delta$ and the associated eigenvector are respectively
\begin{align}
\begin{split}
&\lambda_{+}=\dfrac{1}{2}(p_1-p_2+\sqrt{1-4p_1p_2|\langle \psi^1_{\mathrm{LR}}|\psi^2_{\mathrm{LR}}\rangle|^2},\\
&|+\rangle=\dfrac{1}{\mathcal{N}_+}a(|\psi^1_{\mathrm{LR}}\rangle+|\psi^2_{\mathrm{LR}}\rangle)+b(|\psi^1_{\mathrm{LR}}\rangle-|\psi^2_{\mathrm{RL}}\rangle),
\end{split}
\end{align} 
where $\mathcal{N}_+=\sqrt{\cos^2\theta\sin^2\theta+(\lambda_+(p_1-p_2)\cos^2\theta)^2}$, $a=\sin(\theta)/2$ and $b=(\lambda_+-(p_1-p_2)\cos^2\theta)/(2\sin\theta)$. The optimal POVM, dependent on the spatial overlap of the two spins, is therefore
$\{|+\rangle\langle+|,\mathbb{I}-|+\rangle\langle+|\}$ which gives the probability of error
\begin{align}\label{Perr2}
\begin{split}
P_{\mathrm{err}}&=\dfrac{1}{2}\left(1-\sqrt{1-4p_1p_2|\langle \psi^1_{\mathrm{LR}}|\psi^2_{\mathrm{LR}}\rangle|^2}\right)\\
&=\dfrac{1}{2}-\sqrt{\dfrac{1}{4}-p_1p_2 \left| \dfrac{|lr'|^2e^{i\omega_{\downarrow\uparrow}\phi_{12}}+|l'r|^2e^{i\omega_{\uparrow\downarrow}\phi_{12}}}{\mathcal{N}_{\mathrm{LR}}^2}\right|^2},
\end{split}
\end{align}
where $\phi_{12}=\phi_1-\phi_2$. The probability of error depends on the indistinguishability $(l'r)$ and, for the chosen initial state, is independent of the statistics of particles. The latter can instead play a role for the efficiency of the game starting from a different initial state (see Supplemental Material).

Let us fix $|l|^2=|r'|^2=1/2$: when the two particles spatially overlap with $|l'|^2=|r|^2=\dfrac{1}{2}$, Eq.~\eqref{Perr2} reduces to
\begin{equation}\label{Perrover}
\tilde{P}_{\mathrm{err}}=\dfrac{1}{2}\left(1-\sqrt{1-2p_1p_2\cos^2\left[\left(\dfrac{\omega_{\downarrow\uparrow}-\omega_{\uparrow\downarrow}}{2}\right)\phi_{12}\right]}\right),
\end{equation}
that is zero when $\left(\dfrac{\omega_{\downarrow\uparrow}-\omega_{\uparrow\downarrow}}{2}\right)\phi_{12}=\pi/2$. The behaviour of $P_{\mathrm{err}}$ is displayed in Fig. \ref{Fig1}. 
In particular in Fig.\ref{Fig1} (a) we show, as a function of the phase difference, $\tilde{P}_{\mathrm{err}}$ (blue solid line) and the probability of error $P_{\mathrm{err}}$ for spatially separated particles ($l',r=0$, orange dashed line). This latter case is analogous to that of non-identical particles. Among the two, $\tilde{P}_{\mathrm{err}}$ is always smaller. For non-overlapping spins, being the state incoherent, the best guess is to suppose that the most probable phase $\phi_2$ has been applied. The optimal probability of success is $P_{\mathrm{succ}}=p_2$ and so $P_{\mathrm{err}}=1-P_{\mathrm{succ}}=p_1$ (see orange dashed line in Fig. \ref{Fig1}(a)). In Fig. \ref{Fig1} (b) a contour plot of $P_{\mathrm{err}}$ in terms of $l'$ and $r$ is shown, in the case $\phi_{12}=\pi$. The optimal choice to minimize $P_{\mathrm{err}}$ is to have two overlapping identical spins with $|l'|^2 \sim |r|^2 \sim 1/2$. 

\textit{Conclusions. ---}
In this Letter we have defined the coherence for a system of two identical particles with two-level internal degrees of freedom, in the framework of spatially local operations and classical communication (sLOCC). We have shown that the indistinguishability of identical particles is a source of coherence: while independently prepared distinguishable particles are incoherent under local operations, the analogous configuration with indistinguishable particles can exhibit quantum coherence. In our definition of coherence, one can naturally identify the contribution to coherence exclusively due to the spatial overlap of wavefunctions. In the sLOCC framework, the single- and two-particle incoherent operations known for distinguishable particles are straightforwardly generalized to systems of indistinguishable ones, as we have shown, for instance, in the case of CNOT gate. 

We have then exploited the role of this indistinguishability-enabled coherence in the context of quantum metrology. In particular, we have presented a phase discrimination protocol, for which we can explicitly demonstrate the operational advantage of using indistinguishable rather than distinguishable particles.  Concretely, the coherence 
due to the quantum indistinguishability significantly reduces the 
error probability of guessing the phase using the most general 
measurements.

The fact that coherence due to indistinguishability manifests even for systems of independently prepared identical particles may constitute a practical advantage. In fact, from the one hand it could facilitate the state preparation process and, on the other hand, avoids the typical state fragility linked to the quantum superpositions of composite systems \cite{zurek2009quantum,perez2018endurance}. 

Our results make it clear that spatial overlap of identical particles plays a role not only in the context of entanglement \cite{bose2002indisting,PhysRevLett.88.187903,LoFrancoPRL,chin2019entanglement} but also for another fundamental feature of quantum systems, such as coherence, with an impact in any coherence-based quantum information processes.

\textbf{Acknowledgments.}
AC thanks GA for the hospitality at the University of Nottingham. AW was supported by the Spanish MINECO (project FIS2016-86681-P) with the support of FEDER funds, and the Generalitat de Catalunya (project 2017-SGR-1127).
%\bibliography{references}

\appendix

\section{sLOCC with independently prepared non-identical spins}
Let us consider two independently prepared non-identical ($\mathcal{NI}$) spins, identified by the physical labels A and B, in the two-particle mixed state
\begin{equation}
\rho^{\mathcal{NI}}=\sum_{\sigma,\tau}p_{\sigma\tau}(\ket{\psi \sigma}\bra{\psi \sigma})_{\mathrm{A}}\otimes (\ket{\psi' \tau}\bra{\psi' \tau})_{\mathrm{B}},
\label{NI}
\end{equation}
where $\mathrm{Tr}[\rho^{\mathcal{NI}}]=1$, $\psi$ and $\sigma$ describe the spatial and the spin degrees of freedom of the particle A, while $\psi'$ and $\tau$ of the particle B. Being particles addressable, the framework of spatially local operations and classical communication coincides with the LOCC one. As said in the main text, the operational basis is $\mathcal{B}=\{\ket{\mathrm{L} \sigma}_{\mathrm{A}}\otimes \ket{\mathrm{R} \tau}_{\mathrm{B}}; \sigma,\tau=\downarrow,\uparrow \}$, where $\ket{L}$ and $\ket{R}$ are two states spatially localized in the regions $\mathcal{L}$ and $\mathcal{R}$, respectively. The two regions are chosen such that the probabilities of finding particle A in $\mathcal{L}$ and B in $\mathcal{R}$ are different from zero. Regardless of the spatial overlap of $\ket{\psi}$ and $\ket{\psi'}$, $\rho^{\mathcal{NI}}$ is incoherent, in fact
\begin{align}
\begin{split}
\rho^{\mathcal{NI}}_{\mathrm{LR}}=&\dfrac{1}{\mathcal{N}} \sum_{\sigma\tau}p_{\sigma\tau}\left(\Pi'_{\mathrm{L}} \ket{\psi \sigma}\bra{\psi \sigma} \Pi''_{\mathrm{L}}\right)_{\mathrm{A}}
\otimes \left(\Pi'_{\mathrm{R}}\ket{\psi' \tau}\bra{\psi' \tau} \Pi''_{\mathrm{R}}\right)_{\mathrm{B}}\\
&= \sum_{\sigma\tau}p_{\sigma\tau} (\ket{\mathrm{L}\sigma}\bra{\mathrm{L} \sigma})_{\mathrm{A}}\otimes (\ket{\mathrm{R} \tau}\bra{\mathrm{R} \tau})_{\mathrm{B}},
\end{split}
\end{align}
where $\Pi_{\mathrm{L}}=\sum_{\sigma} (\ket{\mathrm{L}\sigma}\bra{\mathrm{L}\sigma})_{\mathrm{A}}$ and $\Pi_{\mathrm{R}}=\sum_{\sigma} (\ket{\mathrm{R}\sigma}\bra{\mathrm{R}\sigma})_{\mathrm{B}}$. As a result, the mixed state of Eq.~\eqref{NI} for two independently prepared non-identical particles, being diagonal in the chosen basis, is incoherent.
%\section{CNOT gate: incoherent operation for identical particles}
\section{Dependence of the error probability on the particle-statistics}
\begin{figure*}
\centering
\includegraphics[scale=0.45]{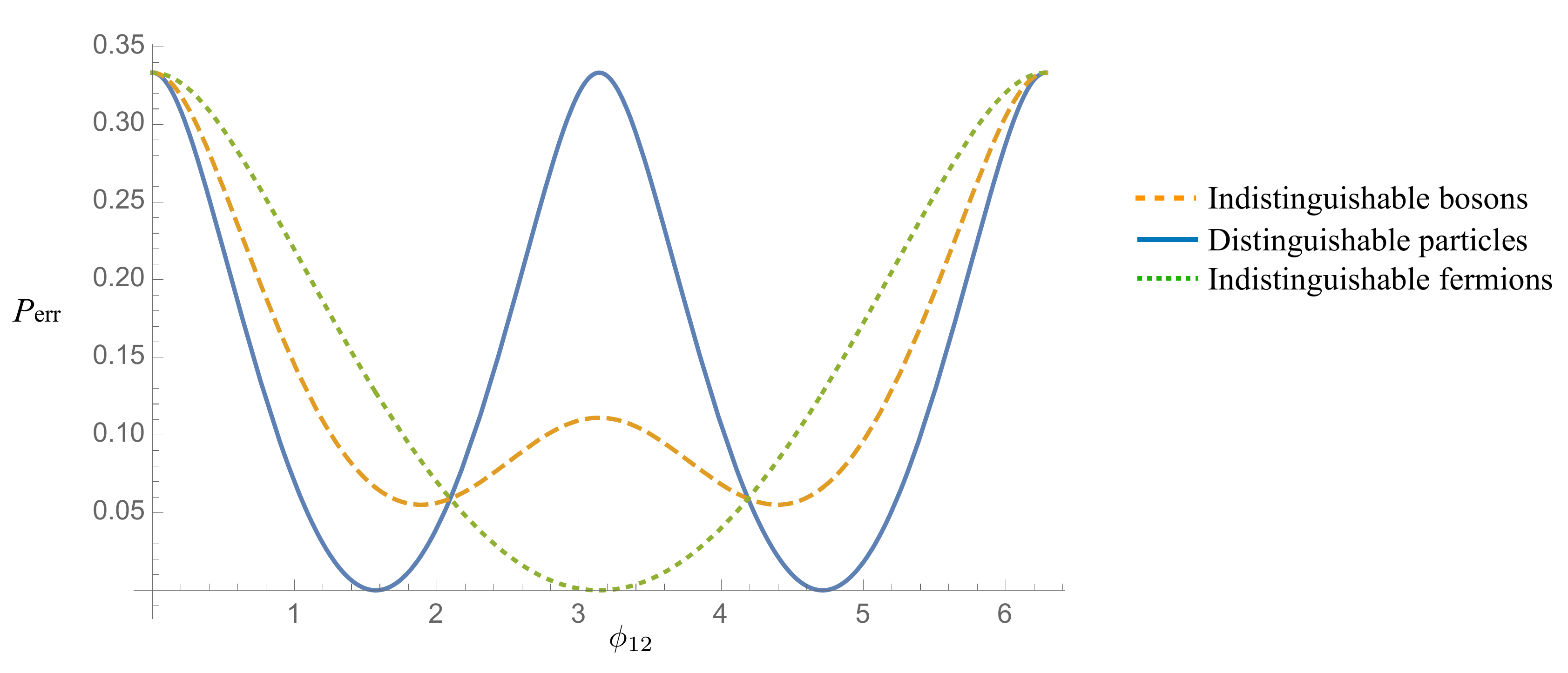}
\caption{Probability of error in function of $\phi_{12}$ with $|l|^2=|r'|^2=1/2$ for identical particles: distinguishable (blue solid line) and indistinguishable with $|l'|^2=|r|^2=1/2$ (orange-dashed line for bosons and green dotted line for fermions), with $a=b$, $\omega_{\downarrow\uparrow}=3$, $\omega_{\uparrow\downarrow}=2$ and $\omega_{\downarrow\downarrow}=1$. %Around $\phi_{12}=\pi$ (in particular from $\phi_{12}=2.0944$ and $\phi_{12}=4.18879$) the coherence due to indistinguishability significantly reduces the probability of error.}
}
\label{sezione}
\end{figure*}

\begin{figure*}
\centering
\includegraphics[scale=0.45]{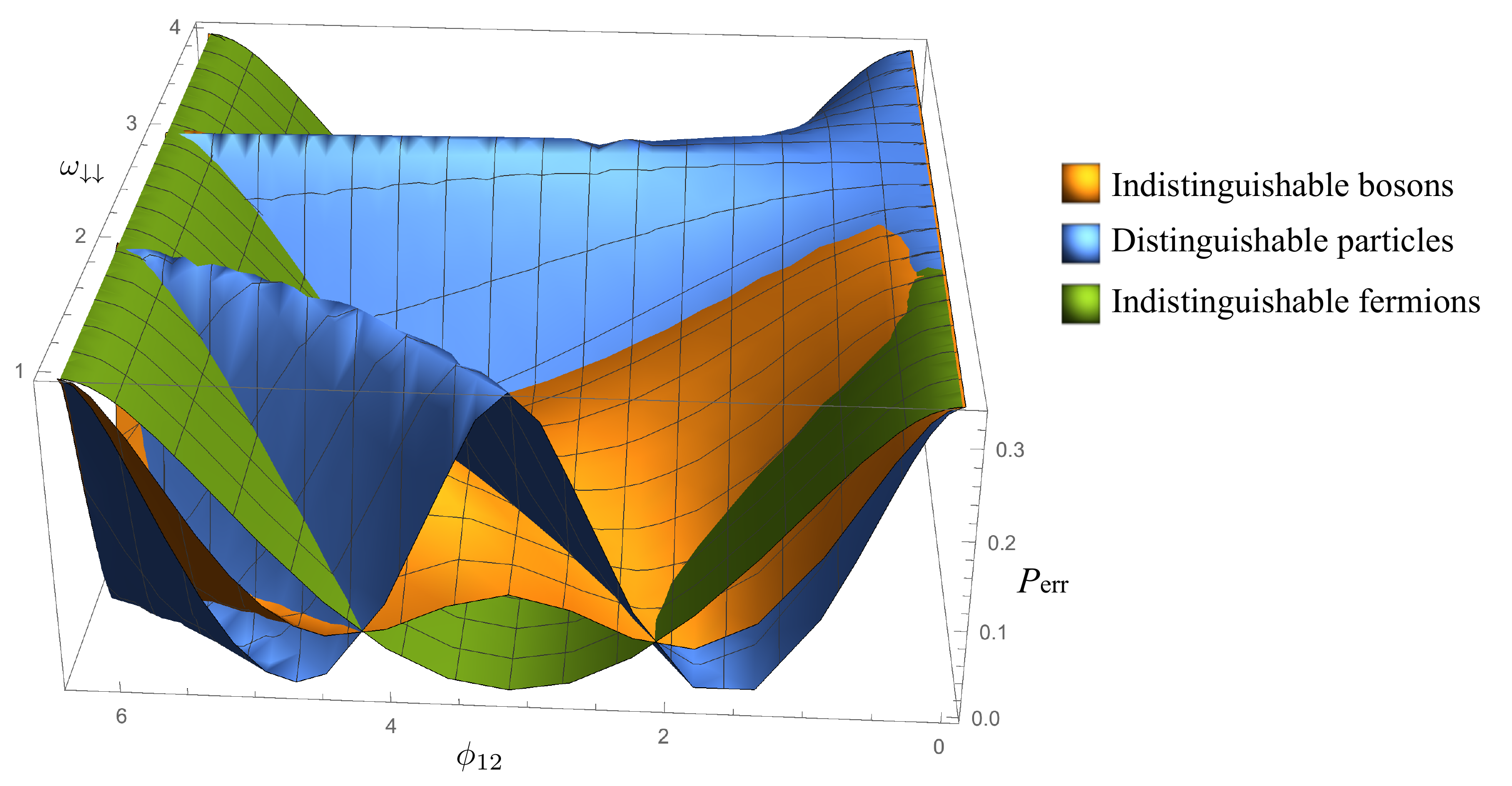}
\caption{3D-plot of the probability of error with $|l|^2=|r'|^2=1/2$ for identical particles: distinguishable (blue) and indistinguishable with $|l'|^2=|r|^2=1/2$ (orange for bosons and green for fermions) in function of $\phi_{12}$ and of $\omega_{\downarrow\downarrow}$, with $|l|^2=|r'|^2=1/2$, $a=b$, $\omega_{\downarrow\uparrow}=3$, $\omega_{\uparrow\downarrow}=2$. }
\label{3dplot}
\end{figure*}

%\begin{figure*}
%\includegraphics[scale=0.5]{suppmat}
%\caption{(a). Probability of error in the discrimination game with bosons for $a=b$, $p_1=1/3$, $|l|^2=|r'|^2=1/2$, $\omega_{\downarrow\downarrow}=1$ and $\omega_{\downarrow\uparrow}-\omega_{\uparrow\downarrow}=1$. The orange dashed line is associated to zero overlap (distinguishable bosons) and the solid blue line for maximum overlap (indistinguishable bosons) with $|l'|^2=|r|^2=1/2$. (b). Contour plot of the probability of error in function of the bosonic spatial overlap with $\phi_{12}=\pi$. Fermions. (c): Probability of error in the discrimination game with fermions with $a=b$, $p_1=1/3$, $|l|^2=|r'|^2=1/2$, $\omega_{\downarrow\downarrow}=1$ and $\omega_{\downarrow\uparrow}-\omega_{\uparrow\downarrow}=1$. The orange dashed line is associated to zero overlap (distinguishable fermions) and the blue to maximum overlap (indistinguishable fermions) with $|l'|^2=|r|^2=1/2$. (d) Contour plot of the probability of error in function of the fermionic overlap with $\phi_{12}=\pi$.}
%\label{Fig2}
%\end{figure*}
The phase discrimination game, within the state of Eq. (3) of the main text, is independent of statistics. However, for a generic state this characteristic is not maintained.
As an example, let us consider the following state
\begin{equation}\label{Psi}
|\Psi\rangle=\dfrac{1}{\mathcal{N}}|\psi \downarrow,\psi's\rangle,
\end{equation}
where $\ket{s}=a\ket{\uparrow}+b\ket{\downarrow}$, with $a,b \in \mathbb{R}$, and $\mathcal{N}$ the normalisation constant. Projecting $\ket{\Psi}$ in the operational subspace spanned by the identical basis $\mathcal{B}^{\mathcal{I}}=\{\ket{\mathrm{L}\sigma,\mathrm{R}\tau}, \ \sigma,\tau=\downarrow,\uparrow\}$, where $\ket{\mathrm{L}}$ and $\ket{\mathrm{R}}$ are localized in two separate regions, we obtain
\begin{equation}\label{PsiRL}
\ket{\Psi_{\mathrm{LR}}}=\dfrac{1}{\mathcal{N}_{\mathrm{LR}}}(\ket{\mathrm{L}\downarrow,\mathrm{R}s'}
+a\eta l'r\ket{\mathrm{L}\uparrow,\mathrm{R}\downarrow}),
\end{equation}
where $\ket{s'}=alr'\ket{\uparrow}+b(lr'+\eta l'r)\ket{\downarrow}$, $\mathcal{N}_{\mathrm{LR}}=\sqrt{a^2(|lr'|^2+|l'r|^2)+b^2|lr'+\eta l'r|^2}$ and $\eta$ distinguishes bosons ($+1$) and fermions ($-1$). The state of Eq.~\eqref{PsiRL} is coherent in $\mathcal{B}^{\mathcal{I}}$ and in the chosen framework of spatially localized measurements, a contribution to the coherence due only to indistinguishability can be identified when particles spatially overlap ($l'r\neq 0$). The latter indistinguishability-enabled contribution plays a statistics-dependent operational role in a phase discrimination game. Let us send $|\Psi_{\mathrm{RL}}\rangle$ in a box in which a unitary transformation $U_k=e^{i\hat{G}\phi_k}$ applies to the state one of the phases $\{\phi_k\}_{k=1,2}$ with a probability $p_k$. Choosing the generator of the transformation as $\hat{G}=\sum_{\sigma\tau}\omega_{\sigma\tau}|\mathrm{L} \ \sigma, \mathrm{R} \ \tau\rangle \langle \mathrm{L} \ \sigma, \mathrm{R} \ \tau |$, with $\sigma,\tau=\downarrow,\uparrow$, we obtain
\begin{align}\label{Psik}
\begin{split}
|\Psi^k_{\mathrm{LR}}\rangle&=U_k |\Psi_{\mathrm{LR}}\rangle\\
&=\dfrac{1}{\mathcal{N}_{\mathrm{LR}}}[ a(lr' e^{i\omega_{\downarrow\uparrow}\phi_k}|\mathrm{L}\downarrow,\mathrm{R}\uparrow\rangle+\eta l'r e^{i\omega_{\uparrow\downarrow}\phi_k}|\mathrm{L}\uparrow,\mathrm{R}\downarrow\rangle)\\
& +b(lr'+\eta l'r)e^{i\omega_{\downarrow\downarrow}\phi_k}|\mathrm{L}\downarrow,\mathrm{R}\downarrow\rangle].
\end{split}
\end{align}
When the state comes out of the box, it is in a classical mixture
\begin{equation}\label{classmixt}
\rho^{out}=\sum_{k=1,2}p_k|\Psi^k_{\mathrm{LR}}\rangle\langle \Psi^k_{\mathrm{LR}}|,
\end{equation}
that represents our ignorance of which phase has been applied to the state. Using Eq. (11) of the main text, the probability of making an error in the discrimination game is now
%\begin{widetext}
%\begin{center}
%\begin{align}\label{Perr2}
%\begin{split}
%P_{\mathrm{err}}&=\dfrac{1}{2}\left(1-\sqrt{1-4p_1p_2|\langle \psi^1_{\mathrm{RL}}|\psi^2_{\mathrm{RL}}\rangle|^2}\right)\\
%&=\dfrac{1}{2}\left(1-\sqrt{1-4p_1p_2 \left| \dfrac{1}{\mathcal{N}_{RL}^2}\left[a^2(|lr'|^2e^{i\omega_{\downarrow\uparrow}\phi_{12}}+a^2|l'r|^2e^{i\omega_{\uparrow\downarrow}\phi_{12}})+b^2|lr'+\eta l'r|^2 e^{i\omega_{\downarrow\downarrow}\phi_{12}}\right]\right|^2}\right),
%\end{split}
%\end{align}
%\end{center}
%\end{widetext}
\begin{widetext}
\begin{center}
\begin{equation}\label{Perr2}
P_{\mathrm{err}}=\dfrac{1}{2}\left(1-\sqrt{1-4p_1p_2 \left| \dfrac{1}{\mathcal{N}_{RL}^2}\left[a^2(|lr'|^2e^{i\omega_{\downarrow\uparrow}\phi_{12}}+|l'r|^2e^{i\omega_{\uparrow\downarrow}\phi_{12}})+b^2|lr'+\eta l'r|^2 e^{i\omega_{\downarrow\downarrow}\phi_{12}}\right]\right|^2}\right),
\end{equation}
\end{center}
\end{widetext}
where $\phi_{12}=\phi_1-\phi_2$. 
%If $a=0$, the initial state is incoherent in any case (see Eq. \ref{Psi}), independently of the spatial overlap, so it is not modified by the unitary operator in the box. Therefore the best strategy for guessing the phase consists of supposing that the most probable phase has been applied. If $b=0$, we obtain the case presented in the main text, in which indistinguishability significantly reduces the probability of error, which is however independent of the statistics.
%On the other hand, if $a$ and $b$ are both different from zero, Eq. \ref{Perr2} depends on the bosonic or fermionic nature of particles according to the value of the parameter $\eta$ that appears in the term depending on $\omega_{\downarrow\downarrow}$. 
From Figs. \ref{sezione} and \ref{3dplot} it is easy to see that for certain ranges of the parameters $\phi_{12}$ and $\omega_{\sigma\tau}$ ($\sigma,\tau=\downarrow,\uparrow$), the probability of error using indistinguishable particles is considerably smaller than the one associated to distinguishable ones. In particular, in these ranges, fermions are more advantageous than bosons.
%For $a=b$, it can be shown that, 
%if $\omega_{\downarrow\uparrow}=\omega_{\uparrow\downarrow}+1=\omega_{\downarrow\downarrow}+2$,
%both overlapping bosons and fermions, with $|l|^2=|r'|^2=|l'|^2=|r|^2=1/2$ (see orange dashed line and green dotted line, respectively, in Fig. \ref{sezione}), give a significant advantage around $\phi_{12}=\pi$ compared with the case of distinguishable particles (blue solid line in Fig. \ref{sezione}). This advantage is greater for fermions, for which the probability of error reaches zero when $\phi_{12}=\pi$. However, there is a range of $\phi_{12}$ in which the probability of error for total overlap is slightly greater than the one associated to zero overlap (see Figs.\ref{sezione}). For generic values of the parameters, it can be more or less advantageous to use indistinguishable particles in the chosen phase discrimination game and this significantly depends on the statistics of particles, as shown in Fig. \ref{3dplot} with a 3D-plot of the error probability in function of $\phi_{12}$ and $\omega_{\downarrow\downarrow}$. The blue surface associated to distinguishable particles intersects with the orange and green ones of indistinguishable bosons and fermions, respectively (see Fig. \ref{3dplot}). For certain ranges of the parameters, it emerges that the probability of error using spatially separate (distinguishable) identical particles is considerably greater than the one associated to spatially overlapping ones. And in particular, in these ranges, fermions are more advantageous than bosons.

\end{document}